# Pulse Radiolysis Studies on Superoxide Reductase from *Treponema pallidum*

Vincent Nivière, ‡* Murielle Lombard, ‡ Marc Fontecave, ‡

and Chantal Houée-Levin §

‡ Laboratoire de Chimie et Biochimie des Centres Redox Biologiques, DBMS-CEA/CNRS/Université Joseph Fourier, 17 Avenue des Martyrs, 38054 Grenoble, Cedex 9, France. § Laboratoire de Chimie Physique, UMR 8000, CNRS/Université Paris-Sud, Bâtiment 350, Centre Universitaire F-91405 Orsay Cedex, France.

* To whom correspondence should be addressed. Telephone: 33-4-38-78-91-09; Fax: 33-4-38-78-91-24; E-mail: vniviere@cea.fr

Footnote

[1] The abbreviations used are: SOD, superoxide dismutase; SOR, superoxide reductase.

ABSTRACT

Superoxide reductases (SORs) are small metalloenzymes, which catalyze reduction of $O_2^-$ to $H_2O_2$. The reaction of the enzyme from *Treponema pallidum* with superoxide was studied by pulse radiolysis methods. The first step is an extremely fast bi-molecular reaction of the ferrous center with $O_2^-$, with a rate constant of $6 \times 10^8$ $M^{-1}$ $s^{-1}$. A first intermediate is formed which is converted to a second one with a slower rate constant of 4800 $s^{-1}$. This latter value is 10 times higher than the corresponding one previously reported in the case of SOR from *Desulfoarculus baarsii*. The reconstituted spectra for the two intermediates are consistent with formation of transient iron-peroxide species.

INTRODUCTION

Superoxide radical ($O_2^-$) is the univalent reduction product of molecular oxygen. It belongs to the group of the so-called toxic oxygen derivatives, which also include hydrogen peroxide and hydroxyl radicals [1]. For years the only enzymatic system known to catalyze the elimination of superoxide was the superoxide dismutase (SOD) [1], which catalyzes dismutation of superoxide radical anions to hydrogen peroxide and molecular oxygen [2]. However, it was recently discovered that biological elimination of $O_2^-$ could also occur by reduction [3, 4], a reaction catalyzed by an enzyme thus named superoxide reductase (SOR) :

$$SOR_{red} + O_2^- + 2\,H^+ \rightarrow H_2O_2 + SOR_{ox}$$

$$SOR_{ox} + 1e^- \rightarrow SOR_{red}$$

---

$$O_2^- + 1e^- + 2\,H^+ \rightarrow H_2O_2$$

SORs are small iron proteins (14 kDa) that contain an unusual mononuclear iron center, consisting of a ferrous iron with square-pyramidal coordination to four nitrogens from histidines as equatorial ligands and one sulfur from a cysteine as the axial ligand, Fe(N-His)$_4$(S-Cys) [5, 6]. This ferrous center is the active site of the enzyme, as it reduces superoxide specifically and very efficiently [3]. The SOR proteins found in some sulfate reducing bacteria, e.g.

*Desulfoarculus baarsii* and *Desulfovibrio vulgaris* contain an additional mononuclear iron center, coordinated by four cysteines in a distorted rubredoxin-type center [3, 5]. However, this iron center is not apparently involved in the reaction and, up to now, its function remains unknown [3]. Very recently, the reactivity of SORs from *Desulfoarculus baarsii* [7] and *Desulfovibrio vulgaris* [8] has been studied in great detail by site-directed mutagenesis and pulse radiolysis studies. The data provided the evidence of reaction intermediates, most likely iron-peroxide species, during the course for the reduction of $O_2^-$ to $H_2O_2$, by the ferrous center.

Recently, we and others have characterized a new type of SOR from the human pathogenic bacteria *Treponema pallidum* [9, 10]. Its primary sequence is homologous to the SORs previously described, except that it is lacking three of the four cysteines involved in the rubredoxin-like iron center. Consequently, we showed that the protein from *Treponema pallidum* lacks this iron center and contains only the mononuclear Fe(N-His)$_4$(S-Cys) iron center. The fact that it exhibits a full SOR activity confirms that the rubredoxin-like iron center is apparently not involved in the reaction.

In this work, because of the presence of only one iron center which makes the enzyme from *T. pallidum* more suitable for studying SOR reaction, we have done pulse radiolysis studies on this protein. The results show significant differences as far as the kinetics of the formation of the reaction intermediates are concerned, with respect to the data recently reported for the SORs from *D. baarsii* [7] and *D. vulgaris* [8].

MATERIALS AND METHODS

Sodium formate and phosphate were of the highest quality available (Prolabo Normatom or Merck Suprapure). Oxygen was delivered by ALPHA GAZ. Its purity is higher than 99.99%. Water was purified using an Elga Maxima system (resistivity 18.2 MΩ). The recombinant SOR from *Treponema pallidum* protein was purified according to the procedure previously described [9]. The ferricyanide oxidized purified protein exhibited a $A_{280nm}/A_{650nm}$ ratio of 9.2. Protein concentration was determined using the value of the molar extinction coefficient of 2300 $M^{-1}$ $cm^{-1}$ at 650 nm for the ferricyanide fully oxidized protein. For the pulse radiolysis experiment, the concentration of the reduced SOR was determined by subtraction of that of the oxidized protein in the solution. The purified SOR was 15% oxidized, and remained stable in this redox state for several hours, in the presence of air or oxygen (1 bar).

Pulse Radiolysis Experiments. Free radicals were generated by the application into an aqueous solution of a 200 ns pulse of high energy electrons, ca. 4 MeV from a linear accelerator located at the Curie Institute, Orsay France [7]. The doses per pulse (2-15 Gy) were calibrated from the absorption of the thiocyanate radical SCN$^{•-}$ obtained by radiolysis of thiocyanate ion solution in N$_2$O-saturated phosphate buffer ([SCN$^-$] =10$^{-2}$ M, 10 mM phosphate, pH 7, G(SCN$^{•-}$)= 0.55 µmol. J$^{-1}$, 472 nm, ε = 7580 M$^{-1}$ cm$^{-1}$). Superoxide radicals were generated during scavenging of radiolytically generated HO$^•$ free radicals by 100 mM formate, in O$_2$ saturated solution, as previously described [7]. This radical is obtained in pure form in less than one microsecond. Samples to be irradiated were made up in 10 mM phosphate buffer, pH 7.0, 100 mM sodium formate, and saturated with pure O$_2$. The doses per pulse were ca. 5 Gy ([O$_2^-$] ≈ 3 µM). The reaction was followed spectrophotometrically between 350 and 750 nm, in a 2 cm path length cuvette designed for pulse radiolysis experiments. Time-dependent absorbance differences were recorded on a digital oscilloscope from two different experiments at two different time scales.

RESULTS AND DISCUSSION

The SOR from *T. pallidum*, 60-130 µM in phosphate buffer pH 7.0, was challenged by 3 µM superoxide, generated radiolytically. O$_2^-$ reacted very rapidly with the iron center and all the absorbances between 350 and 700 nm increased in the microsecond time scale. Fig. 1 shows a representative trace at 575 nm. All the traces between 350 and 700 nm reached a maximum ca. 40 µs after the pulse (shown at 575 nm, Fig. 1). Then the absorbances slowly decayed during *ca.* one millisecond, as shown in Fig. 2 for a representative trace at 550 nm, on the millisecond time scale (up to 1.5 ms). Finally, on a longer time scale, traces at the different wavelengths revealed a further transformation of the iron center to give the final product of the reaction, the ferric iron center (data not shown). But, unfortunately, because of the lamp instability, the kinetics of the reaction could not be further investigated after about 5-10 ms reaction time.

Kinetics were analyzed at all the different wavelengths investigated between 450 and 700 nm. At each wavelength, traces on the microsecond time scale (up to 300 µsec, Fig. 1) could be described as the sum of two exponential processes. The apparent rate constant for the first step of the reaction, resulting in a maximum increase of the absorbance between 500 and 650 nm *ca.* 40 µsec after the pulse, was proportional to protein (reduced form) concentration between 60 and 130 µM (data not shown). The rate constant value, determined from the slope of the straight line $k_{app}$ versus [reduced protein] was found to be $6 \times 10^8$ $M^{-1}$ $s^{-1}$. The second exponential process could be followed at the millisecond time scale to completion (Fig. 2) and was characterized by a rate constant value of 4800 $s^{-1}$, independent of protein concentration at all wavelengths investigated (shown at 550 nm, Fig. 2).

Altogether, these results are consistent with a very fast bi-molecular reaction of SOR with $O_2^{\cdot-}$ ($k = 6 \times 10^8$ $M^{-1}$ $s^{-1}$), leading to the formation of a first reaction intermediate whose concentration is maximum *ca.* 40 μs after the beginning of the reaction. This intermediate then undergoes a transformation to give a second reaction intermediate, which is fully formed 1 msec after the pulse (Scheme 1). The spectra of these two intermediates could be reconstructed from the absorbances at different wavelengths obtained after 40 μs and 1 ms reaction for the first and second intermediate, respectively (Fig. 3). They display an absorption band at 610 and at 670 nm for the first and the second intermediate, respectively. The molar extinction coefficient at these wavelengths are almost identical (5000 $M^{-1}$ $cm^{-1}$), assuming that the reaction of SOR with 3 μM of $O_2^-$ is quantitative.

The formation of two reaction intermediates with comparable spectra was also observed in the case of the SOR from *D. baarsii* [7]. The first intermediate was formed very rapidly with a rate constant of 1.1x10$^9$ $M^{-1}$ $s^{-1}$, which is close to the value determined in the case of *T. pallidum*. On the contrary, the reported rate constant for the formation of the second intermediate was about 10 times slower than in the case of *T. pallidum* (550 $s^{-1}$ compared to 4800 $s^{-1}$, respectively).

In both proteins, the spectra of the intermediates are consistent with the formation of iron-peroxide species (Scheme 1). As a matter of fact, several models Fe-OOH compounds have been found to absorb visible light in the 500-700 nm region [11]. We make the proposal that the first intermediate is an

iron(III)-peroxo species, resulting from the very fast binding of $O_2^-$ to the solvent accessible coordination site of the reduced iron center [5, 6]. The second intermediate may derive from a protonation process of this iron-peroxo species to give an iron(III)-peroxide complex. Finally, the last part of the reaction would correspond to a second protonation process which results in the release of the reaction product $H_2O_2$. Interestingly, the three SOR proteins characterized up to now at the level of reaction intermediates exhibit marked differences in terms of the rate of formation of the second species [7, 8 and this work]. As already mentioned, this rate is 10 times slower for *D. baarsii* as compared to that for *T. pallidum* protein. In the case of the SOR from *D. vulgaris*, the first intermediate was very rapidly generated ($k = 1.5 \times 10^9$ $M^{-1}$ $s^{-1}$), but then underwent a slow transformation (40 $s^{-1}$) to give the reaction products, with no evidence for a second intermediate [8]. The absence of a second intermediate is puzzling, taking into account that the protein from *D. vulgaris* is homologous to those from *D. baarsii* and *T. pallidum* [9]. One explanation might be that the proton donors involved in the second step of the reaction (Scheme 1) are different from one SOR to the other and/or exhibit different pKa values. In the case of the SOR from *D. vulgaris*, the observed greater stability of the first intermediate might result from a combined effect of a lower pKa value for the proton donor than in the case of *T. pallidum* and *D. baarsii* and a pH value slightly larger (pH 7.8) [8] than that used in our experiments (pH 7.0, this work). Further studies are clearly required to characterize the nature of the proton donor in the different SORs.

In conclusion, the SOR from *Treponema pallidum*, with only one iron center, is here shown to proceed through reaction intermediates similar to those observed with the *D. baarsii* and *D. vulgaris* SORs, containing two iron centers. It thus represents an ideal model to study the mechanism of SORs since the presence of the additional rubredoxin-like iron center make the *D. baarsii* and *D. vulgaris* systems more complicated to study, in particular with spectroscopic methods.

FIGURE LEGENDS

Figure 1. Time-dependent on the µsec scale of the absorbance changes at 575 nm during the reaction of the SOR (85 µM) with $O_2^{\cdot -}$ (3 µM) generated by pulse radiolysis. The dashed lines are the best fit assuming two exponential processes.

Figure 2. Time-dependent on the msec scale of the absorbance changes at 550 nm during the reaction of SOR (85 µM) with $O_2^{\cdot -}$ (3 µM) generated by pulse radiolysis. The dashed lines are the best fit assuming two exponential processes.

Figure 3. Reconstituted spectra of the solution during the reaction of SOR (85 µM) with $O_2^{\cdot -}$ (3 µM) generated by pulse radiolysis, 40 µs (●) and 1 ms (♦) after the pulse.

Scheme 1. Formation of the two intermediate species during the reaction of superoxide with SOR from *Treponema pallidum*.

Fig. 1

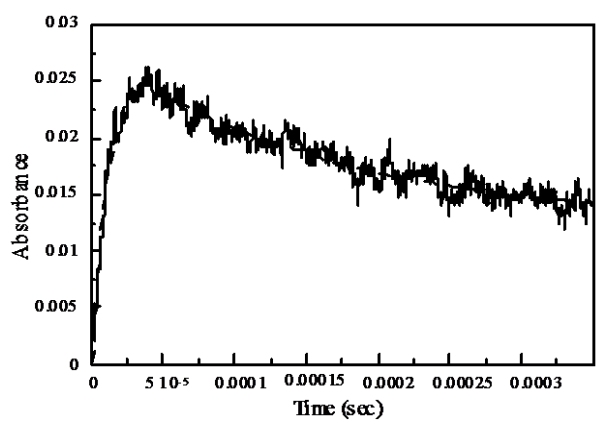

Fig. 2

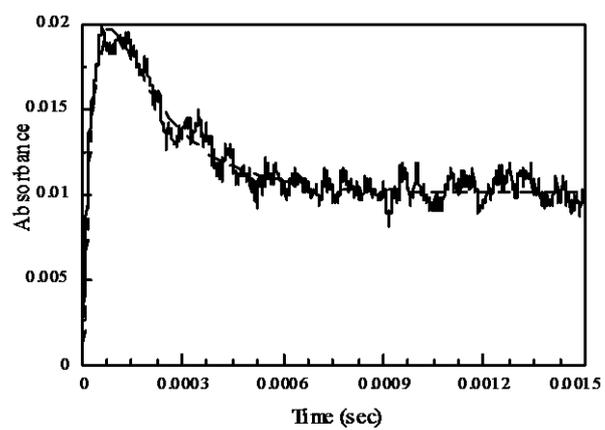

Fig. 3

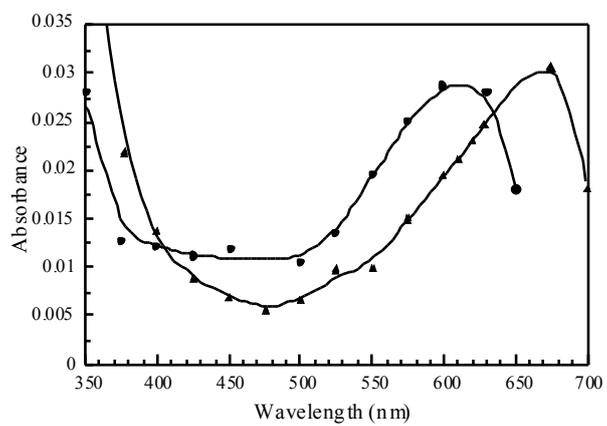

Scheme 1

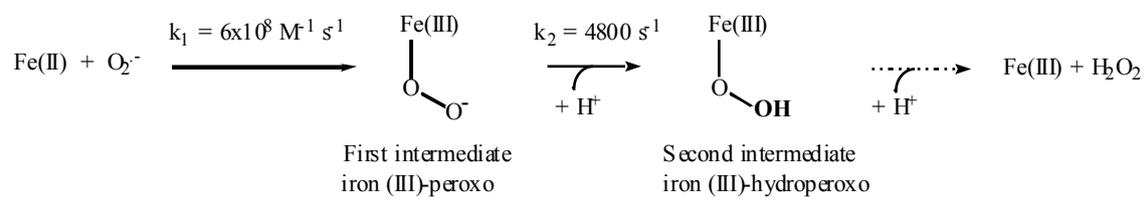